\documentclass[11pt]{article}
\usepackage{amsfonts}
\usepackage{graphicx}
\usepackage{epstopdf}
\usepackage{epsfig}
\usepackage{amssymb}
\usepackage{setspace}
\usepackage{caption}
\usepackage{color}
\usepackage{amsmath}
\usepackage{float}
\usepackage{subcaption}
\newcommand {\be}{\begin{equation}}
	\newcommand {\ee}{\end{equation}}
\newcommand {\bea}{\begin{array}}
	
	\newcommand {\eea}{\end{array}}

\textwidth=6.5in \hoffset=-.75in \voffset=-1in \numberwithin{equation}{section}
\numberwithin{figure}{section}

\begin{document}

	\begin{titlepage}
		%\bigskip \begin{flushright}
			%\\
			%\end{flushright}
			%\maketitle
			\vspace{1cm} 
			\begin{center}
				{\Large \bf {Shadows for Kerr-Sen-Taub-NUT black holes with Manko-Ruiz parameter}}\\
			\end{center}
			\vspace{2cm}
			\begin{center}
				\renewcommand{\thefootnote}{\fnsymbol{footnote}}
				Haryanto M. Siahaan{\footnote{haryanto.siahaan@unpar.ac.id}}\\
				Program Studi Fisika, Universitas Katolik Parahyangan,\\
				Jalan Ciumbuleuit 94, Bandung 40141, Indonesia
				\renewcommand{\thefootnote}{\arabic{footnote}}
			\end{center}
			\vspace{2cm}
			\begin{abstract}

In this paper, we explore the Kerr-Sen-Taub-NUT spacetime with the inclusion of the Manko-Ruiz parameter. We demonstrate the separability of the Hamilton-Jacobi equation for a classical test particle in this spacetime and, using the null Hamilton-Jacobi equation, derive a set of equations that define the black hole shadow’s boundary. Our analysis reveals that the shadow's shape is influenced by variations in the spacetime parameters, including the Hassan-Sen transformation parameter, which corresponds to the black hole’s electric charge, as well as the NUT charge and rotational parameter. Additionally, we highlight the impact of the Manko-Ruiz parameter on the shadow’s deformation, offering new insights into its role in shaping the Kerr-Sen-Taub-NUT spacetime geometry.
				
			\end{abstract}
		\end{titlepage}\onecolumn 
		%%\pacs{04.60.-m,04.62.+v,04.70.-s,04.70.dy,11.25.-w}
		\bigskip 
		
\section{Introduction}\label{sec.intro}
\label{sec:intro}

The Kerr-Sen-Taub-NUT black hole represents an extended solution incorporating both electric charge and the NUT parameter into the family of rotating and charged black hole solutions \cite{Siahaan:2019kbw,Galtsov:1994pd}. Originating from heterotic string theory in four dimensions, this solution provides a compelling framework for examining how additional parameters in Kerr-like spacetimes influence the geometry of black hole shadows. The NUT parameter, in particular, introduces a "gravitomagnetic" monopole moment, resulting in distinctive spacetime properties, such as potential violations of asymptotic flatness \cite{Griffiths:2009dfa,Aliev:2008wv,Aliev:2007fy,Lynden-Bell:1996dpw,Page:1978hdy}. Moreover, the Kerr-Sen-Taub-NUT solution captures the unique characteristics of the Taub-NUT spacetime, including closed timelike curves and the Misner string, which can significantly impact the perceived shape and size of the black hole shadow \cite{Zhang:2021pvx,Wu:2023eml}.

Recent studies underscore the continued interest in black hole solutions arising in the low-energy limit of heterotic string theory. For instance, the accelerating Kerr-Taub-NUT spacetime in this framework is presented in \cite{Siahaan:2024ljt}, while the thermodynamic topology of Kerr-Sen black holes using Rényi statistics is discussed in \cite{Zhang:2023svu}. Additionally, the heat engine efficiency of Kerr-Sen-AdS black holes is explored in \cite{Roy:2023qqy}, and the ultraspinning Kerr-Sen-AdS$_4$ black holes are analyzed in \cite{Wu:2020cgf}. Further investigations include photon emissions near the horizons of Kerr-Sen black holes \cite{Zhang:2020pay} and the weak cosmic censorship conjecture for Kerr-Sen black holes \cite{Gwak:2019rcz}. These studies collectively emphasize the rich phenomenology and theoretical significance of Kerr-Sen and Kerr-Sen-Taub-NUT spacetimes.

The incorporation of the Manko-Ruiz parameter further extends the black hole model by modifying the Kerr metric through a complex parameterization technique, enabling exploration of spacetime configurations with even more intricate symmetries. Originally developed to describe rotating solutions with mass quadrupole moments in general relativity, the Manko-Ruiz parameter provides a systematic way to assess deviations from spherical symmetry in a charged, rotating black hole. This parameter thus offers a versatile approach to studying how higher-order multipole moments affect observable features, such as the black hole shadow, and serves as a useful extension when examining possible non-Kerr deviations \cite{Manko:2005nm}. With these parameters, the Kerr-Sen-Taub-NUT model facilitates detailed examinations of shadow distortions due to charge, rotation, NUT charge, and the Manko-Ruiz parameter as well.

In recent years, black hole shadows have attracted significant attention as a crucial observational feature revealing details about spacetime geometry in the vicinity of compact astrophysical objects. Black hole shadow formation arises from the bending and capture of light near a black hole, yielding a dark region against a brighter background, observed as an apparent "shadow." This shadow provides a direct observational link to the nature of gravity, offering potential insights into both general relativity and alternative theories of gravity. Building on the first-ever image of a black hole shadow captured by the Event Horizon Telescope (EHT), there has been a surge of interest in studying shadows in various black hole metrics, aiming to explore deviations from Kerr geometry and the implications of additional parameters \cite{EventHorizonTelescope:2019dse,EventHorizonTelescope:2022wkp}.

Recent studies investigating black hole shadows in contexts similar to the approach adopted in this paper can be summarized as follows. In \cite{Kumar:2024cnh}, the authors analyzed the shadow of a newly discovered black hole solution in Einstein gravity, incorporating Ayon–Beato–Garcia nonlinear electrodynamics and a cloud of strings. Meanwhile, the shadow of rotating black holes in bumblebee gravity was examined in \cite{Islam:2024sph}. An intriguing contribution is presented in \cite{Ghasemi-Nodehi:2024bcv}, where a constraint on the NUT charge was proposed based on shadow observations of Sgr A*. Additionally, gravitational lensing and shadow effects around a Schwarzschild-like black hole in metric-affine bumblebee gravity were explored in \cite{Gao:2024ejs}. The study in \cite{Liu:2024soc} discussed light rings and the shadow of static black holes within the Hamiltonian constraint approach to effective quantum gravity. Furthermore, investigations into the shadows of renormalization group-improved rotating black holes, $\mathbb{S}\mathbb{T}\mathbb{U}$ black holes, higher-dimensional MOG dark compact objects, and Kerr-MOG-(A)dS black holes were conducted in \cite{Sanchez:2024sdm, Sekhmani:2024dhc, Nozari:2024jiz, Liu:2024lbi}. Finally, \cite{Kostaros:2024vbn} reported fractal signatures of non-Kerr spacetimes in the shadows of light-ring bifurcations. Collectively, these recent studies underscore the ongoing significance of black hole shadow research, which remains a central focus of the black hole physics community.

In this paper, we investigate the shadow edge of a Kerr-Sen-Taub-NUT black hole incorporating the Manko-Ruiz parameter. A prior study \cite{Melvernaldo:2022hdv} examined the separability of the Hamilton-Jacobi equation for the Kerr-Sen-Taub-NUT black hole without the Manko-Ruiz parameter. By computing and plotting shadow contours, we aim to elucidate how the various parameters influence the shadow's shape. This analysis offers valuable insights into observational signatures that could distinguish such exotic black holes from the standard Kerr black hole.

The structure of the paper is as follows: Section \ref{sec.KSTN} introduces the Kerr-Sen-Taub-NUT-Manko-Ruiz black hole metric and its essential properties. In Section \ref{sec.HJ}, we explore the Hamilton-Jacobi equation in this spacetime, followed by a detailed study of the black hole shadow in Section \ref{sec.Shadow}. Finally, we provide our conclusions. Throughout this paper, we adopt natural units where $c=G=\hbar=1$.

\section{Kerr-Sen-Taub-NUT solution with Manko-Ruiz parameter}\label{sec.KSTN}

Kerr-Sen-Taub-NUT (KSTN) solution is obtained by employing the Hassan-Sen transformation to the Kerr-Taub-NUT (KTN) solution as the seed. Since we are interested in studying the Kerr-Sen-Taub-NUT spacetime with Manko-Ruiz (MR) parameter, let us consider the KTN spacetime equipped with MR parameter $C$, namely \cite{Manko:2005nm}
\[
{\rm{d}}{\tilde s}^2  =  - \frac{{{\Delta}  - a^2 \sin^2\theta }}{\Sigma }{\rm{d}}t^2  + \frac{{2\left( {\chi {\Delta}  - a\sin^2\theta \left( {a\chi  + \Sigma } \right)} \right)}}{\Sigma }{\rm{d}}t{\rm{d}}\phi  + \Sigma \left( {\frac{{{\rm{d}}r^2 }}{{{\Delta} }} + {{\rm{d}}\theta^2 }} \right)
\]
\be \label{eq.KTN}
+ \frac{{\left( {\left( {a\chi  + \Sigma } \right)^2 \sin^2\theta  - \chi ^2 {\Delta} } \right)}}{\Sigma }{\rm{d}}\phi ^2 
\ee 
where $\Sigma  = r^2  + \left( {l + a\cos\theta} \right)^2 $, ${\Delta}  = r^2  - 2mr + a^2  - l^2 $, and $\chi  = a\sin^2\theta  - 2l\left( {\cos\theta + C} \right)$. It is known that the KTN solution solves the Einstein's vacuum equations, and the constant $C$ in the metric functions denotes the type of conic singularity in the spacetime. 

In principle, any metric that belongs to a vacuum solution of Einstein equations $R_{\mu\nu}=0$ which possesses the stationary and axial-symmetric properties should be a good seed in Hassan-Sen transformation to construct an exact solution in the low energy limit of heterotic string theory. In \cite{Siahaan:2018qcw}, the Hassan-Sen transformation in the low energy limit of heterotic string theory is presented in a more direct approach where each of the field content $\{G_{\mu\nu}, A_\mu, B_{\mu \nu}, \Phi\}$ is expressed as functions of the seed metric $\tilde{g}_{\mu\nu}$. Here, $G_{\mu\nu}$ is the metric in the string frame, $A_\mu$ is the Maxwell vector field, $\Phi$ is the dilaton, and $B_{\mu\nu}$ is the second-rank antisymmetric tensor field (also known as the Kalb-Ramond field), which couples to the three-index field strength $H_{\mu\nu\lambda}$ defined later.

In terms of seed metric ${\tilde g}_{\mu\nu}$, the string-frame metric solution can be written as
\be 
{\rm d}s_{\rm string}^2 = \frac{{{\tilde g}_{tt} }}{\Lambda^2 }\left( {{\rm d}t + \frac{{{\tilde g}_{t\phi } }}{{{\tilde g}_{tt} }}\left( {1 + s^2 } \right){\rm d}\phi } \right)^2 +{\tilde g}_{rr}{\rm d}r^2 +{\tilde g}_{\theta\theta}{\rm d}\theta^2 +  \left({\tilde g}_{\phi \phi }  - \frac{{{\tilde g}_{t\phi }^2 }}{{{\tilde g}_{tt} }}\right) {\rm d}\phi
\ee
where $\Lambda = 1+s^2\left(1+{\tilde g}_{tt}\right)$. The vector field can be expressed as
\be 
A_\mu  dx^\mu   = \frac{{2s\sqrt {1 + s^2 } }}{\Lambda }\left( {\left( {1 + {\tilde g}_{tt} } \right){\rm d}t + {\tilde g}_{t\phi } {\rm d}\phi } \right)\,,
\ee
whereas the dilaton field reads $\Phi  =  - \ln \Lambda$, and the non-vanishing components of second-rank tensor field reads
\be
B_{t\phi }  =  - B_{\phi t}  = \frac{{s^2 {\tilde g}_{t\phi } }}{\Lambda }\,.
\ee
In the equations above, $s$ is the transformation parameter which in the generic Kerr-Sen solution corresponds to the black hole electric charge\footnote{In Sen's original work \cite{Sen:1992ua}, the parameter \( s \) is defined as \( s = \sinh\frac{\alpha}{2} \), where \( \alpha \) is an arbitrary constant.}. The case $s=0$ associates to the identity transformation, and imposing this condition to the KSNTN solution yields the generic Kerr black hole of the vacuum Einstein system. The fields $\left\{G_{\mu\nu}, A_\mu, B_{\mu \nu}, \Phi \right\}$ obey the equations of motion
\be
R_{\alpha \beta }  - \frac{1}{2}G_{\alpha \beta } R + \nabla _\alpha  \nabla _\beta  \Phi  - G_{\mu \nu } \left[ {\nabla ^2 \Phi  - \frac{{\left( {\nabla \Phi } \right)^2 }}{2}} \right] = \frac{1}{4}\left[ {F_{\alpha \mu } F_\beta ^\mu   + H_{\alpha \mu \nu } H_\beta ^{\mu \nu }  - \frac{{F^2 }}{4}G_{\alpha \beta }  - \frac{{H^2 }}{6}G_{\alpha \beta } } \right]\,,
\ee
\be 
\nabla _\alpha  F^{\alpha \beta }  = F^{\alpha \beta } \nabla _\alpha  \Phi  + \frac{1}{2}F^{\mu \nu } H_{\beta \mu \nu } \,,
\ee 
\be 
\nabla _\alpha  H^{\alpha \mu \nu }  = H^{\alpha \mu \nu } \nabla _\alpha  \Phi \,,
\ee 
and
\be 
\left( {\nabla \Phi } \right)^2  - 2\nabla ^2 \Phi  = R - \frac{{F^2 }}{8} - \frac{{H^2 }}{{12}}\,,
\ee 
where $F^2  = F_{\mu \nu } F^{\mu \nu } $, $
H^2  = H_{\mu \nu \alpha } H^{\mu \nu \alpha} $, and the tensor field with three indices $H_{\mu \nu \lambda }$ is defined as
\be 
H_{\mu \nu \lambda }  = \partial _\mu  B_{\nu \lambda }  + \partial _\lambda  B_{\mu \nu }  + \partial _\nu  B_{\lambda \mu }  - \frac{1}{4}\left( {A_\mu  F_{\nu \lambda }  + A_\lambda  F_{\mu \nu }  + A_\nu  F_{\lambda \mu } } \right)\,.
\ee 

The Einstein-frame metric solution for KSTN spacetime can be obtained by the formula $
{\rm{d}}s^{\rm{2}}  = \exp \left( { - \Phi } \right){\rm{d}}s_{\rm string}^2$. Accordingly, the non-vanishing Einstein-framed metric tensor components can be expressed as
\be \label{eq.gtt}
g_{tt}  = \frac{{a^2 \sin ^2 \theta  - {\Delta} }}{\Xi }\,,
\ee 
\be 
g_{t\phi }  = \frac{{\left( {1 + s^2 } \right)\left( {{\Delta} \chi  - a\sin ^2 \theta \left( {a\chi  + \Sigma } \right)} \right)}}{\Xi }\,,
\ee 
\be 
g_{rr}  = \frac{\Xi }{{{\Delta} }}=\frac{g_{\theta\theta}}{{\Delta}} \,,
\ee 
and
\[
g_{\phi \phi }  = \frac{\Sigma}{\Xi^2 }\left[ {\Sigma ^2 \sin ^2 \theta \left( {1 + s^2 } \right)^2  - 2\Sigma \left( {1 + s^2 } \right)\sin ^2 \theta \left( {s^2 {\Delta}  - a\chi \left( {1 + s^2 } \right)} \right)} \right.
\]
\be  \label{eq.gpp}
\left. { + \left( {{\Delta}  - a^2 \sin ^2 \theta } \right)\left( {s^4 {\Delta} \sin ^2 \theta  - \left( {1 + s^2 } \right)^2 \chi ^2 } \right)} \right]\,,
\ee 
where $\Xi = \Sigma \left(1+s^2\right)+s^2 \left(a^2 \sin^2\theta - {\Delta} \right)$. In the KSTN metric, the black hole mass $M$ and electric charge $Q$ are related to the parameters $m$ and $s$ as follows \cite{Sen:1992ua}
\be 
m = M - \frac{Q^2}{2M}, \quad s = \frac{Q}{\sqrt{2M^2 - Q^2}}\,.
\ee
For a simpler notation and to maintain consistency with previous works, we retain $m$ and $s$ in our equations. This choice simplifies the metric functions and avoids unnecessary algebraic complexity when handling field equations and related calculations. However, whenever $m$ appears, it should be understood in terms of the black hole mass $M$ via the relation above. In the numerical evaluations of the black hole shadow below, we will express the black hole mass and electric charge explicitly as $M$ and $Q$.

\begin{figure}[H]
	\centering
	\includegraphics[scale=0.4]{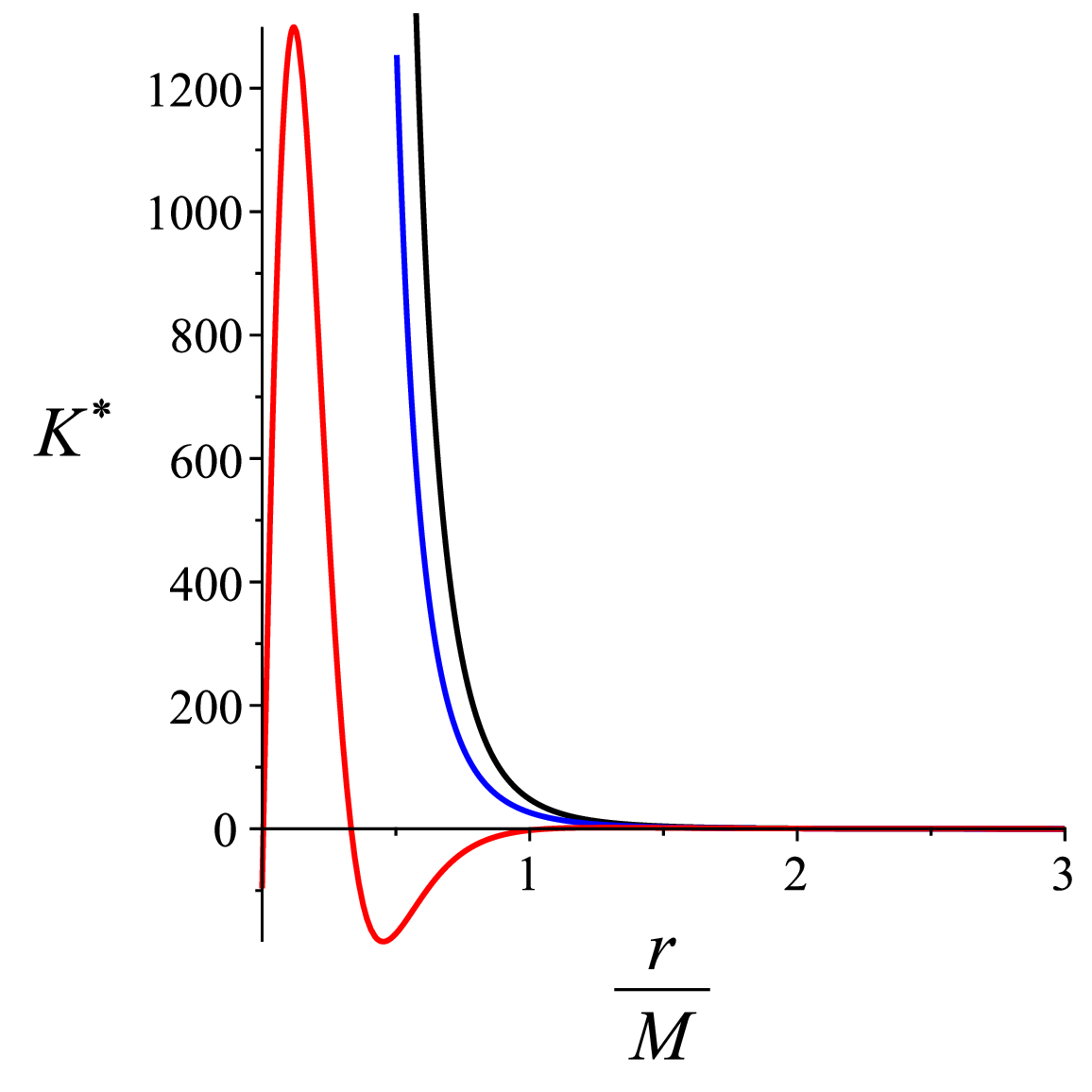}
	\caption{The plots above are equatorial Kretchmann scalar for some particular considerations. The black curve describes the Kerr spacetime with $a=0.1~M$. The blue one is for Kerr-Sen for $Q=0.5~M$ and $a=0.1~M$. The red curve represents the Kerr-Sen-Taub-NUT case with $l=0.5~M$, $a=0.1~M$, and $Q=0.5$.}\label{fig.KK}
\end{figure}

To distinguish the KSTN spacetime with its null NUT parameter case, let us study the Kretchmann scalar associated to the spacetime. Interestingly, the Manko-Ruiz parameter $C$ does not contribute to this quantity. From Fig. \ref{fig.KK}, we can observe the typical property of NUTty spacetime where the physical singularity at origin is absent. 
Similar to the seed solution in Eq. (\ref{eq.KTN}), the Kerr-Sen-Taub-NUT metric above also possesses the same symmetries generated by $\zeta^\mu_{\left(t\right)} = (1,0,0,0)$ and $\zeta^\mu_{\left(\phi\right)} = (0,0,0,1)$ Killing vectors. These vectors obey the Killing equation $\nabla _\mu  \zeta _\nu   + \nabla _\nu  \zeta _\mu   = 0$.

\section{Hamilton-Jacobi equation for a test object}\label{sec.HJ}

To study the motion of a classical test object in curved spacetime, let us consider the Lagrangian
\be 
{\cal L} = \frac{1}{2}g_{\mu \nu } \frac{{dx^\mu  }}{{d\sigma }}\frac{{dx^\nu  }}{{d\sigma }}\,,
\ee 
where $\sigma$ is some affine parameters. The stationary and axial-symmetric nature of the spacetime allow the existence of conserved quantities $E$ and $L$, namely the test object's energy and angular momentum, respectively. They are given by
\be 
- E = g_{t\mu } \frac{{dx^\mu  }}{{d\sigma }} =  g_{tt}\frac{{dt}}{{d\sigma }} + g_{t\phi}\frac{{d\phi }}{{d\sigma }} \,,
\ee 
and
\be 
L = g_{\phi \mu } \frac{{dx^\mu  }}{{d\sigma }} = g_{t\phi}\frac{{dt}}{{d\sigma }} + g_{\phi\phi}\frac{{d\phi }}{{d\sigma }}\,,
\ee 
where the metric tensor components are given in eqs. (\ref{eq.gtt}) - (\ref{eq.gpp}). Accordingly, the dependence of $t$ and $\phi$ on the affine parameter $\lambda$, which describes the null geodesics, is given by
\be 
\Xi \frac{{{\rm{d}}\phi }}{{{\rm{d}}\lambda }} = \left( {1 + s^2 } \right)\left( {\frac{{a^2 \chi  + \Sigma a}}{{{\Delta} }} - \frac{\chi }{{\sin ^2 \theta }}} \right)E + \left( {\frac{1}{{\sin ^2 \theta }} - \frac{{a^2 }}{{{\Delta} }}} \right)L \,,
\ee 
and
\[
\Xi \frac{{{\rm{d}}t}}{{{\rm{d}}\lambda}} = \left[ {\left( {1 + s^2 } \right)^2 \left( {\frac{{\left( {a\chi  + \Sigma } \right)^2 }}{{{\Delta} }} - \frac{{\chi ^2 }}{{\sin ^2 \theta }}} \right) + s^4 \left( {{\Delta}  - 2\Sigma  - a^2 \sin ^2 \theta } \right) - 2s^2 \Sigma } \right]E
\]
\be 
+ \left( {1 + s^2 } \right)\left( {\frac{\chi }{{\sin ^2 \theta }} - \frac{{\chi a^2  + a\Sigma }}{{{\Delta} }}} \right)L\,.
\ee 

Furthermore, due to the stationary and axial symmetric properties of the spacetime under consideration, we can take an ansatz 
\be 
S =  - Et + L\phi  + S_r \left( r \right) + S_\theta \left( \theta \right)
\ee
for the action in Hamilton-Jacobi equation of a test object $
\partial _\mu  S\partial ^\mu  S = \delta$. The null case is given by $\delta=0$, whereas the timelike case is described by $\delta=-1$.  Accordingly, the Hamilton-Jacobi equation can be expressed as
\[
\left( \frac{{d S_\theta }}{{d\theta }}\right)^2 + {\Delta} \left(\frac{{d S_r }}{{dr }}\right)^2  
- \left[ {s^4 \left( {{\Delta}  - 2\Sigma  - a^2 \sin ^2 \theta } \right) - 2s^2 \Sigma  + \left( {1 + s^2 } \right)^2 \left( {\frac{{Y^2 }}{{{\Delta} }} - \frac{{\chi ^2 }}{{\sin ^2 \theta }}} \right)} \right]E^2 
\]
\be 
+ 2\left( {1 + s^2 } \right)\left( {\frac{{aY }}{{{\Delta} }} - \frac{{\chi }}{{\sin ^2 \theta }}} \right)EL + \left( {\frac{1}{{\sin ^2 \theta }} - \frac{{a^2 }}{{{\Delta} }}} \right)L^2 -\delta \Xi =0 \,,
\ee 
where $Y=r^2+l^2+a^2-2alC$. The fact where $\Xi$ is separable between terms that depend on $r$ and $\theta$ yields the separability of this Hamilton-Jacobi equation in the radial and angular dependences. 

Since we are interested in studying the Shadow of KSTN black hole, let us focus on the null version of Hamilton-Jacobi equation discussed previously. In such consideration, the last equation can be separated into the radial equation
\be 
{\Delta} \left( {\frac{{dS_r }}{{dr}}} \right)^2  + \left[ {2r^2 s^2 \left( {1 + s^2 } \right) - {\Delta} s^4 } \right]E^2  - \frac{{\left( {aL - EY\left( {1 + s^2 } \right)} \right)^2 }}{{{\Delta} }} =  - K
\ee 
and the angular one
\be 
\left( {\frac{{dS_\theta  }}{{d\theta }}} \right)^2  + \left[ {a^2 s^2 \sin ^2 \theta  + 2\left( {1 + s^2 } \right)\left( {l + a\cos \theta } \right)^2 } \right]s^2 E^2 + \frac{\left(L-E\chi \left(1+s^2\right)\right)^2}{\sin^2\theta} = K \,.
\ee 
Moreover, by recalling that
\be 
g_{\alpha \beta } \frac{{dx^\beta  }}{{d\lambda }} = \frac{{\partial S}}{{\partial x^\alpha  }}\,,
\ee 
we can have
\be 
\Xi \frac{{dr}}{{d\lambda }} = \sqrt {\cal R} \,,
\ee 
and
\be 
\Xi \frac{{d\theta }}{{d\lambda }} = \sqrt \Theta  \,,
\ee 
where
\be \label{eq.R}
{\cal R} =   {\left( {aL - EY\left( {1 + s^2 } \right)} \right)^2 }   - \left[K + \left(L-aE\right)^2 + \left({2r^2 s^2 \left( {1 + s^2 } \right) - {\Delta} s^4 }\right) E^2\right] {\Delta} \,,
\ee 
and
\be 
\Theta = K + \left(L-aE\right)^2 - \left[ {a^2 s^2 \sin ^2 \theta  + 2\left( {1 + s^2 } \right)\left( {l + a\cos \theta } \right)^2 } \right]s^2 E^2 - \frac{\left(L-E\chi \left(1+s^2\right)\right)^2}{\sin^2\theta} \,.
\ee 
From these equations, we can understand that the circular motion associates to the condition ${\cal R}=0$, and equatorial one corresponds to the $\Theta=0$ consideration.

Now by defining a new affine parameter $\lambda$ where
\be 
\Xi \frac{{dx^\mu  }}{{d\lambda }} = \frac{{dx^\mu  }}{{d\sigma }} \equiv \dot x^\mu \,, 
\ee 
now we can have
\[
{\dot t} = \left[ {\left( {1 + s^2 } \right)^2 \left( {\frac{{\left( {a\chi  + \Sigma } \right)^2 }}{{{\Delta} }} - \frac{{\chi ^2 }}{{\sin ^2 \theta }}} \right) + s^4 \left( {{\Delta}  - 2\Sigma  - a^2 \sin ^2 \theta } \right) - 2s^2 \Sigma } \right]E
\]
\be \label{eq.rdot}
+ \left( {1 + s^2 } \right)\left( {\frac{\chi }{{\sin ^2 \theta }} - \frac{{\chi a^2  + a\Sigma }}{{{\Delta} }}} \right)L \,,
\ee 
\be 
{\dot r} = \sqrt{{\cal R}} \,,
\ee 
\be 
{\dot \theta} = \sqrt{\Theta} \,,
\ee 
and
\be\label{eq.phidot} 
{\dot \phi} = \left( {1 + s^2 } \right)\left( {\frac{{a^2 \chi  + \Sigma a}}{{{\Delta} }} - \frac{\chi }{{\sin ^2 \theta }}} \right)E + \left( {\frac{1}{{\sin ^2 \theta }} - \frac{{a^2 }}{{{\Delta} }}} \right)L \,.
\ee
The results above are required to study the motions of test objects, including the light-like which leads to the following discussion of black hole shadow.

The propagation of light in the KSTN spacetime obeys eqs. (\ref{eq.rdot}) - (\ref{eq.phidot}). Accordingly, we can introduce  ${\bar L}=L/{E}$ and ${\bar K}=K/{E}^{2}$ to describe the motion of photon near the black hole. At this stage, it is useful to apply equation (\ref{eq.R}) to outline the shape of the shadow formed by the black hole, which is determined by the conditions
\be 
{\cal R} = 0~~~~{\rm and}~~~~ \frac{{\rm d {\cal R}}}{{\rm d}r}=0 \,.
\ee 
These equations give us
\be 
{\bar L} = \frac{{{\cal Y}\Delta ' - \Delta {\cal Y}' - \Delta  W }}{{a\Delta '}}
\ee 
and
\[
{\bar K} = \frac{1}{{\left( {a\Delta '} \right)^2 }}\left\{ {\left[ {2\left( {{\cal Y} - a^2 } \right)W + d{\cal Y}'{\cal Y} + Z'\Delta  - 2a^2 {\cal Y}' - a^2 Z'} \right]\Delta \Delta ' - \left[ {a^2 Z + \left( {{\cal Y} - a^2 } \right)^2 } \right]\left( {\Delta '} \right)^2 } \right.
\]
\be 
\left. { - 2\Delta {\cal Y}'\left( {W + {\cal Y}'} \right)\left( {\Delta  - a^2 } \right)} \right\}\,,
\ee 
where in equations above we have $W = 2 \left(r+ms^2\right)$, ${\cal Y} = Y \left(1+s^2\right)$, and $Z = 2s^2 r^2 \left(1+s^2\right) - s^4 \Delta$. It is understood that the primed notation stands for the derivative with respect to radius $r$.
The radius of photon's spherical orbit $r_p$ is obtained by solving ${\cal R}=0$ and $\frac{{d{\cal R}}}{{dr}} = 0$, simultaneously, as they describe the null geodesics with $\dot r=0$ and $\ddot r=0$. 

\section{Black hole shadows}\label{sec.Shadow}

Having demonstrated the separability of the Hamilton-Jacobi equation for null objects in the KSTN spacetime in the previous section, we now move on to explore the black hole shadow in this spacetime. To study the shadow of KSTN black hole, we adopt the approach outlined in \cite{Cunha:2016bpi} in analyzing the shadow's edge. Just like the seed solution, namely the KTN metric with MR parameter, the KSTN solution discussed in this paper possesses the Killing symmetries generated by the $\partial_t$ and $\partial_\phi$ vectors. Now, to denote the observer basis, let us consider the basis vectors $\left\{ {\hat e_{\left( t \right)} ,\hat e_{\left( r \right)} ,\hat e_{\left( \theta  \right)} ,\hat e_{\left( \phi  \right)} } \right\}$ expanded in the coordinate basis $
\left\{ {\partial _t ,\partial _r ,\partial _\theta  ,\partial _\phi  } \right\}$ as \cite{Cunha:2016bpi}
\be\label{eq.CoeffAB}
{\hat e}_{\left( t \right)}  = {\zeta}^t \partial _t  + {\zeta}^\phi  \partial _\phi  ~~,~~{\hat e}_{\left( \phi \right)}  = {\xi}^\phi  \partial _\phi ~~,~~
{\hat e}_{\left( \theta \right)}  = {\xi}^\theta  \partial _\theta ~~,~~{\hat e}_{\left( r \right)}  = {\xi}^r  \partial _r \,,
\ee 
where $\xi^{r}$, $\xi^{\theta}$, $\xi^{\phi}$, $\zeta^{t}$, $\zeta^{\phi}$ are some real coefficient functions. The normalization of observer basis follows that of Minkowski, namely 
\be \label{eq.norm}
{\hat e}_{\left( a \right)}  \cdot {\hat e}_{\left( b \right)}  = \eta _{\left( a \right)\left( b \right)}  \,,
\ee 
where 
\be 
\eta _{\left( t \right)\left( t \right)} = -\eta _{\left( r \right)\left( r \right)}  = -\eta _{\left( \theta  \right)\left( \theta  \right)}  = -\eta _{\left( \phi  \right)\left( \phi  \right)}   = -1 \,,
\ee 
and $\eta _{\left( a \right)\left( b \right)} =0$ for $a \ne b$. 
Accordingly, one can consider 
\[
\xi ^r  = \frac{1}{{\sqrt {g_{rr} } }}~~,~~\xi ^\theta   = \frac{1}{{\sqrt {g_{\theta \theta } } }}~~,~~\xi ^\phi   = \frac{1}{{\sqrt {g_{\phi \phi } } }} \,,
\]
and
\[
\zeta ^t  = \sqrt {\frac{{g_{\phi \phi } }}{{g_{t\phi }^2  - g_{tt} g_{\phi \phi } }}}~~,~~ \zeta ^\phi   =  - \frac{{g_{t\phi } }}{{g_{\phi \phi } }}\sqrt {\frac{{g_{\phi \phi } }}{{g_{t\phi }^2  - g_{tt} g_{\phi \phi } }}} \,, 
\]
in constructing the observer basis ${\hat e}_{\left(a\right)}$ in terms of coordinate ones.

The locally measured momentum $p^{\left(a\right)}$ is obtained by projecting the four momentum  $p_\mu$ onto the observer basis. For example, the measured energy by the observer would be
\be 
p^{\left(t\right)} = - e^\mu_{\left(t\right)} p_\mu = -\left({\zeta}^t p_t + {\zeta}^\phi p_\phi\right)\,,
\ee 
where the negative sign corresponds to the normalization we consider in eq. (\ref{eq.norm}). Note that $p_t = -E$ and $p_\phi = L$ are two constants of motion for a test object in the axial symmetric and stationary spacetime just like the one we consider in this paper. Therefore, by keeping in mind the normalization (\ref{eq.norm}), we can have
\be \label{eq.pt}
p^{\left( t \right)}  = E{\zeta}^t  - L{\zeta}^\phi  \,,
\ee 
\be \label{eq.pr}
p^{\left( r \right)}  = e_{\left( r \right)}^\mu  p_\mu   = {\xi}^r p_r  = \frac{{p_r }}{{\sqrt {g_{rr} } }} \,,
\ee
\be \label{eq.ptheta}
p^{\left( \theta  \right)}  = e_{\left( \theta  \right)}^\mu  p_\mu   = {\xi}^\theta  p_\theta   = \frac{{p_\theta  }}{{\sqrt {g_{\theta \theta } } }} \,,
\ee 
and
\be \label{eq.pphi}
p^{\left( \phi  \right)}  = e_{\left( \phi  \right)}^\mu  p_\mu   = {\xi}^\phi  p_\phi   = \frac{L}{{\sqrt {g_{\phi \phi } } }} \,.
\ee
Furthermore, the photon linear momentum ${\bf p}$ is constructed by the space-like components of $p^{\mu}$, where the following relation holds
\be 
{\bf p} \cdot {\bf p} = 
\sum\limits_{a,b = r,\theta ,\phi } {\eta _{\left( a \right)\left( b \right)} p^{\left( a \right)} p^{\left( b \right)} } \,.
\ee 
Hence, by using the angular coordinate $\left(\alpha,\beta\right)$, each momentum components can be expressed as
\be 
p^{\left( r \right)}  = \left| {\rm{\bf p}} \right|\cos \alpha \cos \beta ~~,~~p^{\left( \phi  \right)}  = \left| {\rm{\bf p}} \right|\cos \alpha \sin \beta ~~,~~p^{\left( r \right)}  = \left| {\rm{\bf p}} \right|\sin \alpha \,.
\ee 
For the null object, we understand that $p^{\left(t\right)} = \left| {\rm{\bf p}} \right|$. 

Now we can proceed to work out the analysis of black hole shadow using the quantities discussed above. In what follows, we consider an observer in a ZAMO frame positioned at radial and latitudinal coordinates $r_o$ and $\theta_o$, respectively. Since we consider an observer facing the black hole, hence the radial momentum of the photon must be positive, namely $p^{\left(r\right)} \ge 0$. Accordingly, the angular coordinate $\left(\alpha,\beta \right)$ can be expressed as
\be 
\alpha  = \sin ^{ - 1} \left( {\frac{{p^{\left( \theta  \right)} }}{{p^{\left( t \right)} }}} \right)~~,~~\beta  = \tan ^{ - 1} \left( {\frac{{p^{\left( \phi  \right)} }}{{p^{\left( r \right)} }}} \right) \,.
\ee 
Using the momentum components in eqs. (\ref{eq.pt}) - (\ref{eq.pphi}), the explicit expressions for the angles above are given by
\be 
\sin \alpha  = \left. \frac{{\left\{ {\left( {f_2 \cos ^2 \theta  + f_1 \cos \theta  + f_0 } \right)\left( {c_2 \chi ^2  + c_1 \chi  + c_0 } \right)} \right\}^{1/2} }}{{c_2 \chi ^2  + c_1 \chi  + c_0  + \bar L\left( {1 + s^2 } \right)\left( {\chi \Delta  - a\sin ^2 \theta \left( {a\chi  + \Sigma } \right)} \right)}} \right|_{r_o ,\theta _o } \,,
\ee 
and
\be 
\tan \beta  = \left. \frac{{\bar L\Xi \sqrt \Delta  }}{{\left[ {\left( {c_2 \chi ^2  + c_1 \chi  + c_0 } \right)\left( {s^4 \Delta ^2  - \left( {\left( {\bar L - a} \right)^2  + 2s^2 r^2 \left( {1 + s^2 } \right) + \bar K} \right)\Delta  + \left( {a\bar L - Y\left( {1 + s^2 } \right)} \right)^2 } \right)} \right]}} \right|_{r_o ,\theta _o } 
\ee
where
\[
f_2  =  - 2s^2 a^2 \sin ^2 \theta \left( {1 + s^2 } \right) \,,
\]
\[
f_1  =  - 4\sin ^2 \theta s^2 la\left( {1 + s^2 } \right) \,,
\]
\[
f_0  =  - a^2 s^2 \sin ^2 \theta  + \left( {\left( {\bar L - a} \right)^2  + \bar K - 2s^2 l^2 \left( {1 + s^2 } \right)} \right)\sin \theta  - \left( {\bar L - \chi \left( {1 + s^2 } \right)} \right) \,,
\]
\[
c_2  = \left( {1 + s^2 } \right)\left( {a^2 \sin ^2 \theta  - \Delta } \right) \,,
\]
\[
c_1  = 2\Sigma a\sin ^2 \theta \left( {1 + s^2 } \right)^2  \,,
\]
and
\[
c_0  = \sin ^2 \theta \left( {\left( {1 + s^2 } \right)^2 \Sigma ^2  - 2s^2 \Delta \left( {1 + s^2 } \right) + \Delta s^4 \left( {\Delta  - a^2 \sin ^2 \theta } \right)} \right) \,.
\]

\begin{figure}[H]
	\centering
	\begin{subfigure}{.5\textwidth}
		\centering
		\includegraphics[width=.8\linewidth]{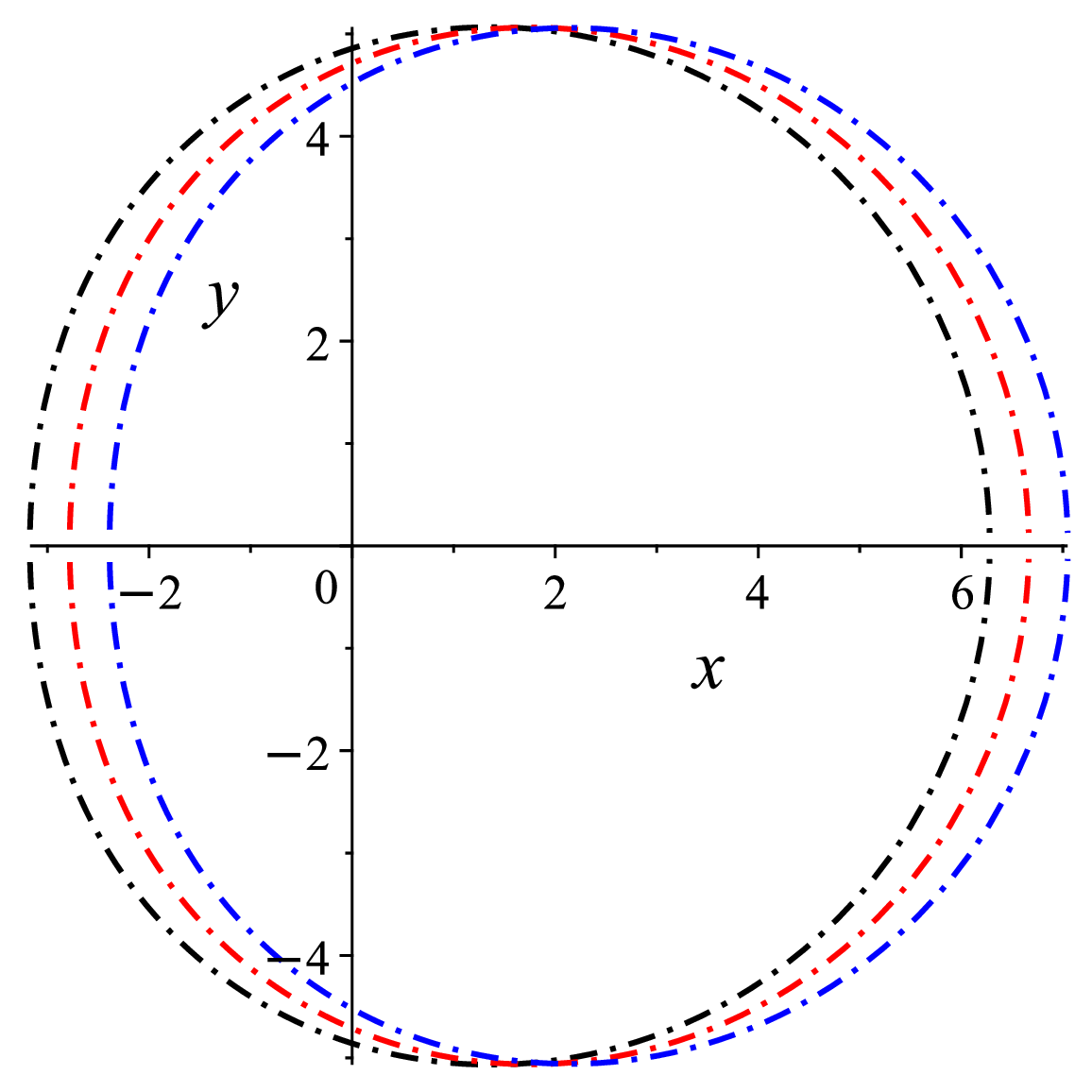}
		\caption{}
		\label{fig:shadowCm1C0Cp1l02}
	\end{subfigure}%
	\begin{subfigure}{.5\textwidth}
		\centering
		\includegraphics[width=.8\linewidth]{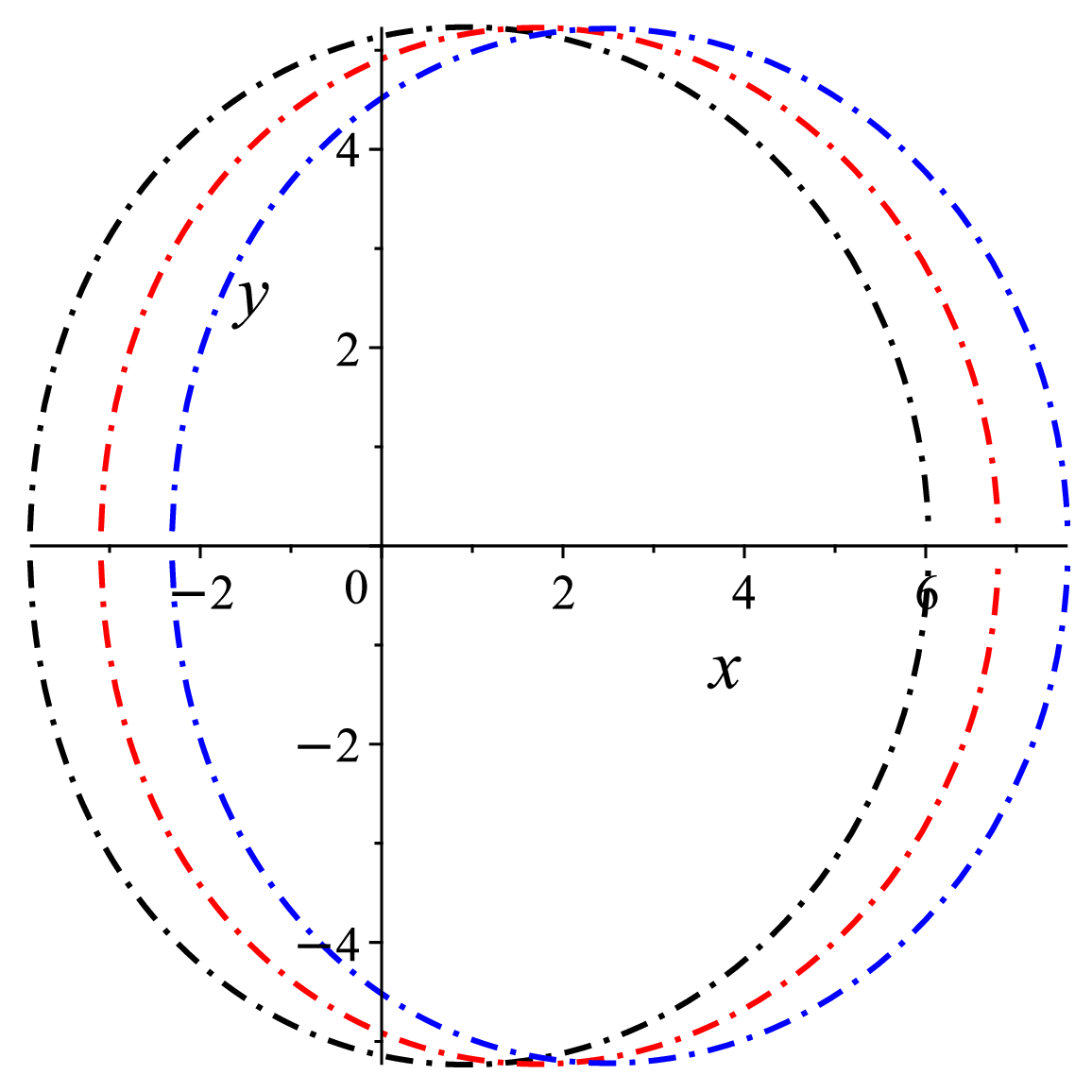}
		\caption{}
		\label{fig:shadowCm1C0Cp1l04}
	\end{subfigure}
	\caption{Shadow edge plots for the Kerr-Sen-Taub-NUT black hole observed at the equator with $r_o = 30~M$. The black, red, and blue curves correspond to $C=-1$, $C=0$, and $C=1$, respectively. In getting those plots, we have considered $a=0.9~M$ and $Q=0.2M$. Plots in Fig. \ref{fig:shadowCm1C0Cp1l02} is for $l=0.2~M$ case, whereas Fig. \ref{fig:shadowCm1C0Cp1l04} describes the $l=0.4~M$ consideration.}
	\label{fig:P3}
\end{figure}

Moreover, we consider the Cartesian coordinates $\left(x,y\right)$ related to the plane of observer's sight which are given by
\be 
x = -r_o \beta ~~,~~ y=r_o \alpha \,,
\ee 
where $r_o$ is the radial position of static observer. These Cartesian coordinates depend on $\theta_o$, namely the inclination angle between the rotational axis and the observer. In Fig. \ref{fig:P3}, we provide some numerical evaluations for the shadow's edge plots which vary for different MR parameters and NUT charge considerations. A typical property for such shadow presents, namely it is symmetric with respect to the $x$-axis. From this figure, we also learn how different consideration of MR parameter, i.e. the location of string singularity in the spacetime, would modify the observed shadow on the observer's plane of observation. In particular, from Fig. \ref{fig:P3}, we also learn how the variation of NUT parameter contribute to the change of shadow's edge, i.e. it changes in size and deformation from the circular shape.  

\begin{figure}[H]
	\centering
	\begin{subfigure}{.5\textwidth}
		\centering
		\includegraphics[width=.8\linewidth]{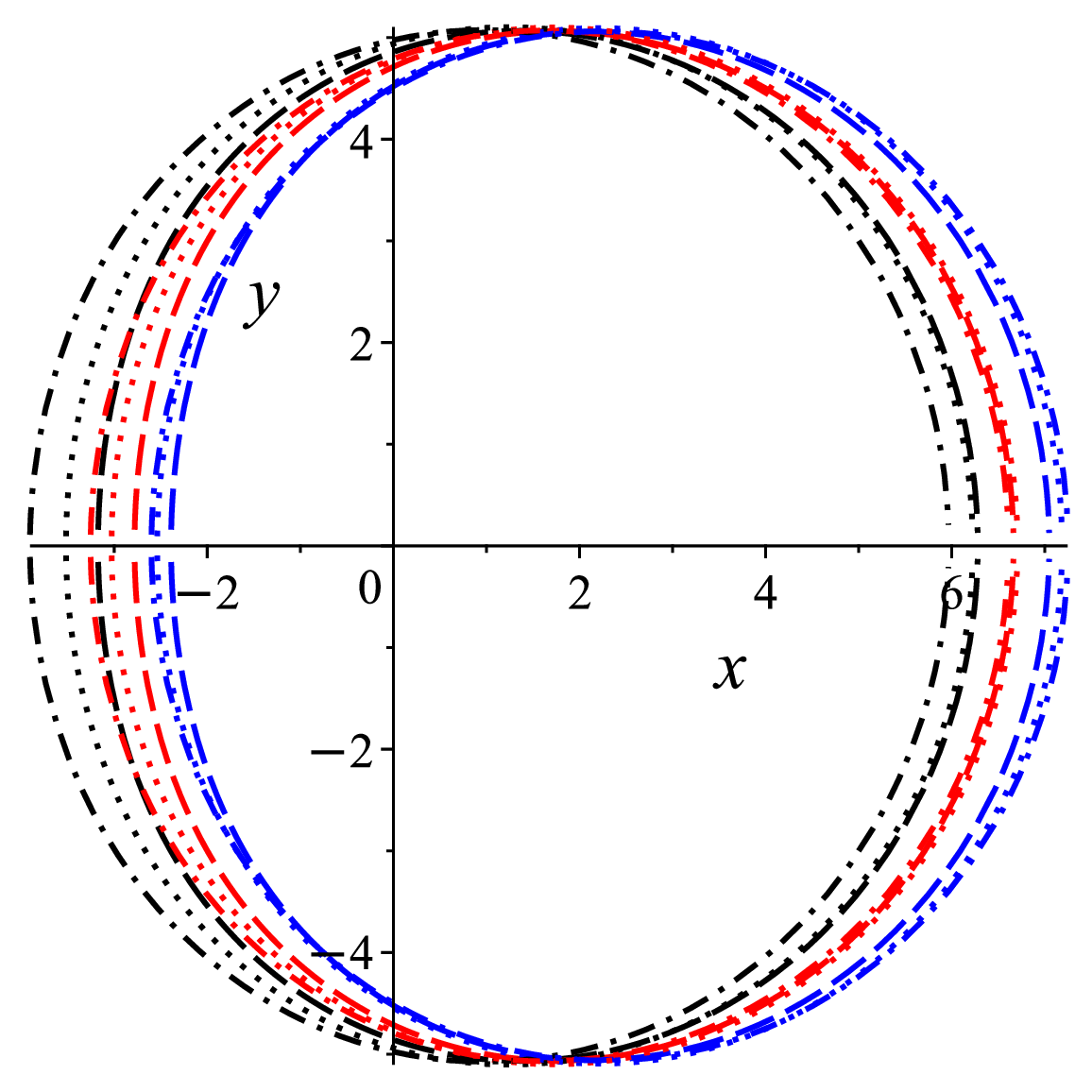}
		\caption{}
		\label{fig:shadowThCm1C0Cp1l02}
	\end{subfigure}%
	\begin{subfigure}{.5\textwidth}
		\centering
		\includegraphics[width=.8\linewidth]{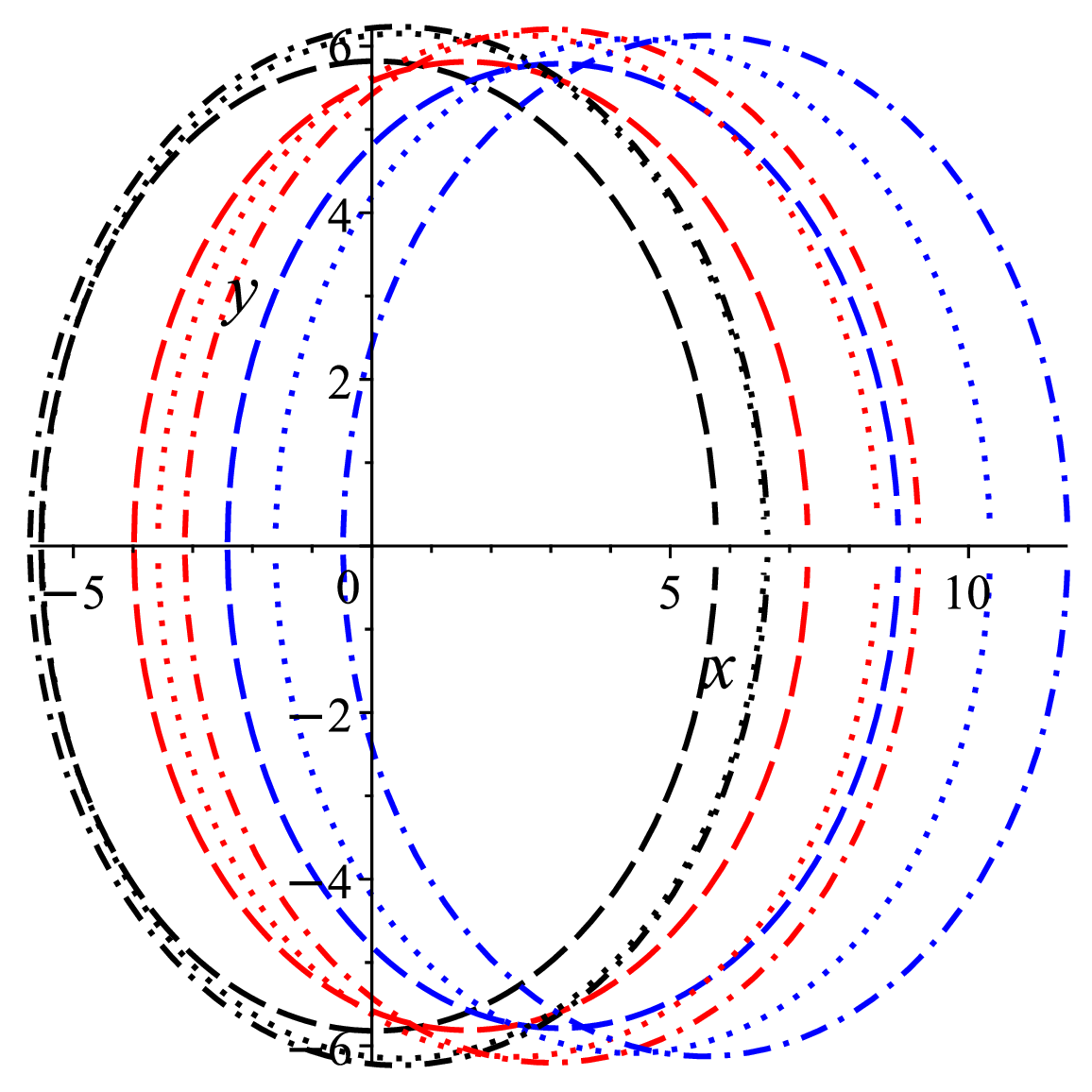}
		\caption{}
		\label{fig:shadowThCm1C0Cp1l04}
	\end{subfigure}
	\caption{Plots of shadow's edge for Kerr-Sen-Taub-NUT black hole observed at several inclination angles of the observer at $r_o=30~M$. Again, black, red, and blue curves correspond to the cases of $C=-1$, $C=0$, and $C=1$, respectively. In getting those plots, we have considered $a=0.9~M$ and $Q=0.2~M$. The dashed, dots, and dashed-dots curves correspond to the $\theta_o = \pi/2$, $\theta_o = \arcsin 0.8$, and $\theta_o = \arcsin 0.6$ cases, respectively. Plots in Fig. \ref{fig:shadowThCm1C0Cp1l02} correspond to the $l=0.2~M$ case, whereas Fig. \ref{fig:shadowThCm1C0Cp1l04} represents the $l=0.8~M$ consideration.}\label{fig:P4}
\end{figure}

For the variations of shadow's edge for some observer's inclination angle $\theta_o$, we provide Fig. \ref{fig:P4}. The general behavior of the curves is the center of shadow moves the positive $x$ direction as the MR parameter increases. The same behavior is also found for the variation of $\theta_o$ for the considered angles, namely $\theta_o = \pi/2$ or equatorial, $\theta_o = \arcsin 0.8$, and $\theta_o = \arcsin 0.6$. An interesting feature is also observed in Fig. \ref{fig:P4}, related to different values of the NUT parameter. For $l=0.2~M$ in Fig. \ref{fig:shadowThCm1C0Cp1l02}, the size of shadow for several MR parameters and different $\theta_o$ is roughly similar, or we cannot find from the figure some significant discrepancy for the size.  However, in Fig. \ref{fig:shadowThCm1C0Cp1l04}, we can see some different size of shadow as the inclination angle $\theta_o$ varies. 

\begin{figure}[H]\centering 
	\includegraphics[scale=0.4]{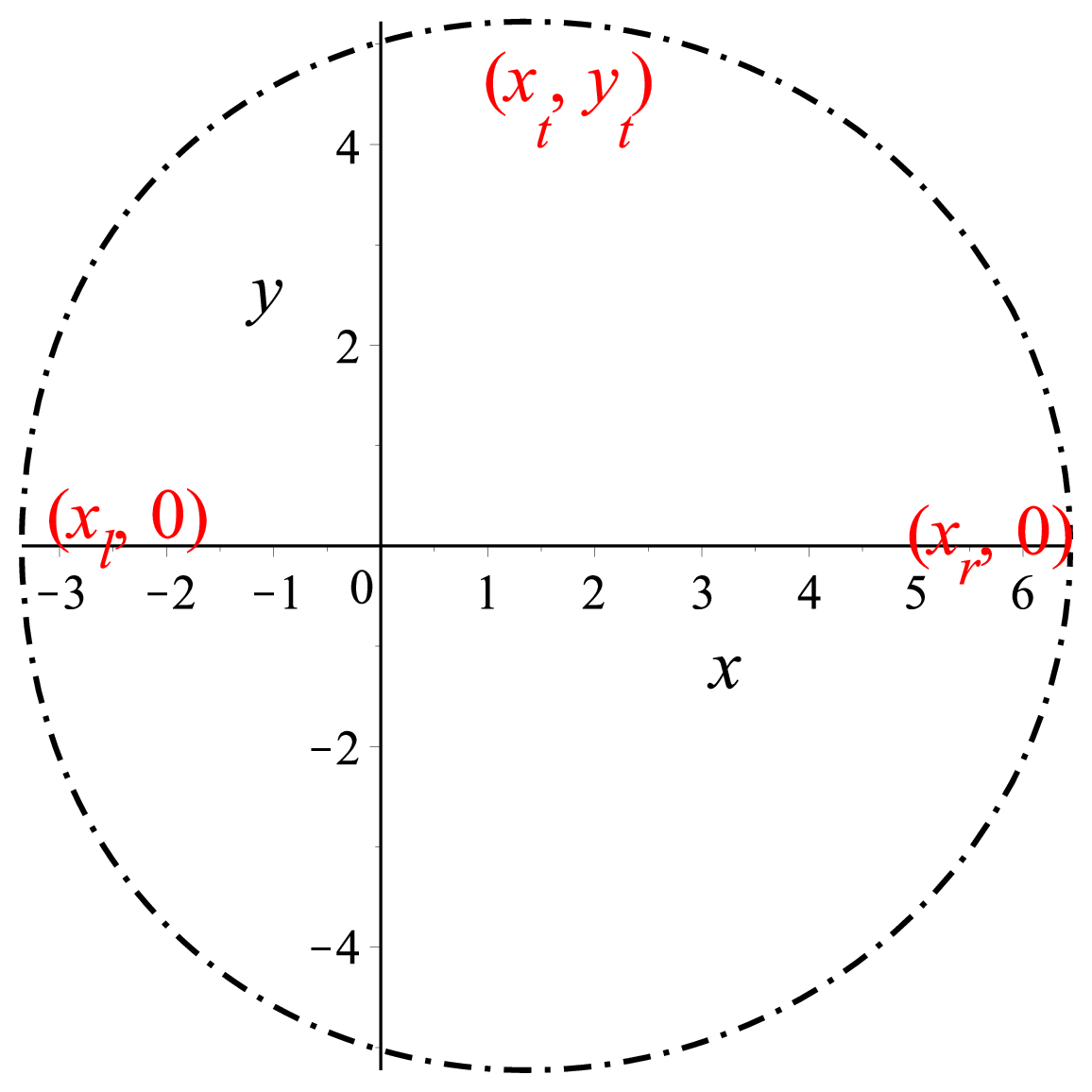}\caption{Cartoon for Cartesian points to compute the observables $R_s$ and $\delta_s$. The top point in the curve is denoted by $(x_t,y_t)$, whereas $(x_l,0)$ and $(x_r,0)$ are the points where the curve intersects horizontal axis.}\label{fig.cartoon}
\end{figure}

Nevertheless, from figs. \ref{fig:P3} and \ref{fig:P4}, it is not too easy to quantify how different the shadow as some parameters of the spacetime are varied. They look quite similar, and in the form of expected one for a shadow of a black hole. To further analyze the shadow, we use the observables introduced by Hioki and Maeda \cite{Hioki:2009na}, namely, the radius
\be \label{eq.Rs}
R_s  = \frac{{\left( {x_r  - x_t } \right)^2  + y_t^2 }}{{2\left( {x_r  - x_t } \right)}} \,,
\ee 
and the deformation
\be \label{eq.ds}
\delta _s  = \frac{{x_l  - x_r  + 2R_s }}{{R_s }}\,.
\ee 
The cartoon illustrating points used in these observables is given in Fig. \ref{fig.cartoon}. 

\begin{figure}[H]
	\centering
	\begin{subfigure}{.5\textwidth}
		\centering
		\includegraphics[width=.8\linewidth]{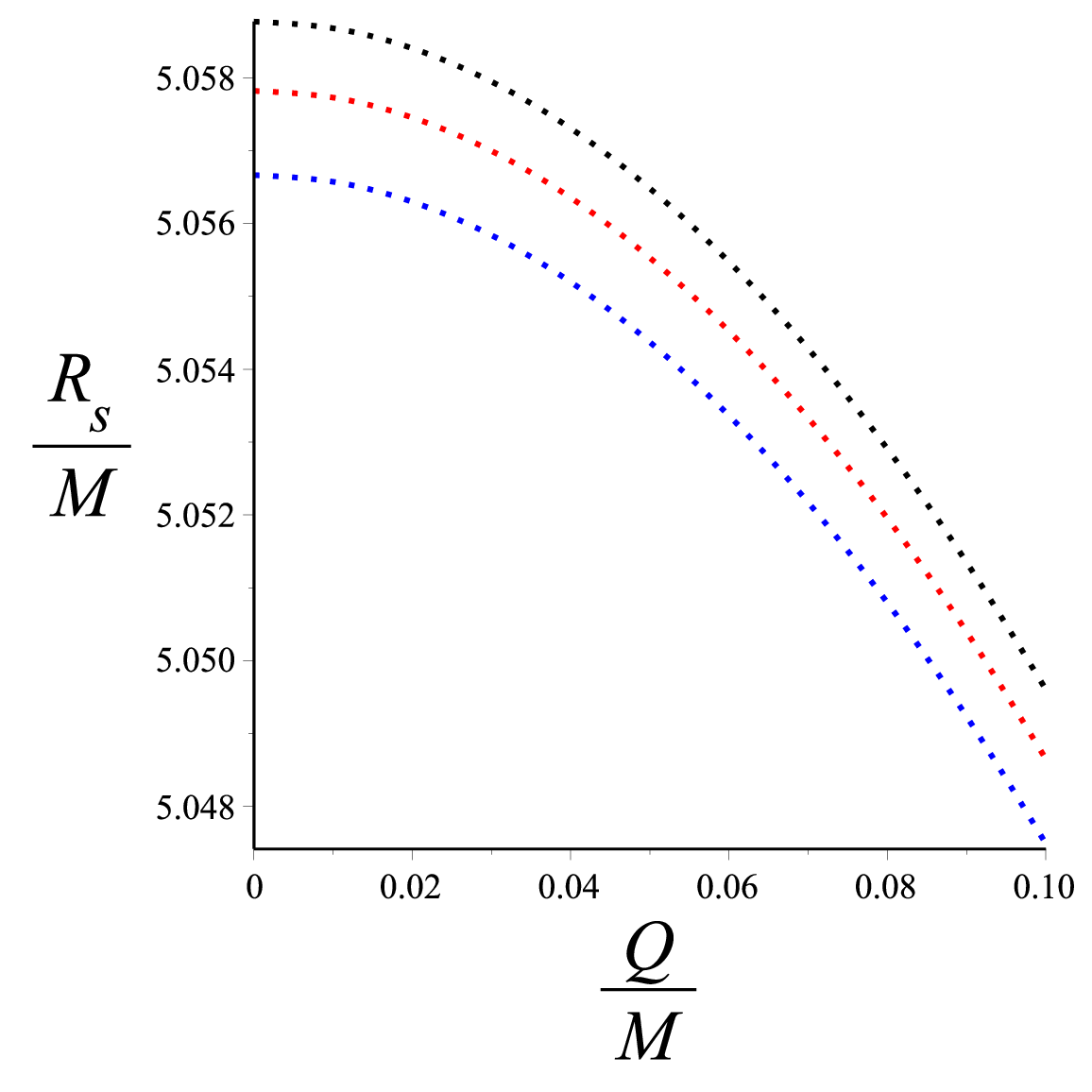}
		\caption{}
		\label{fig:Rsl01}
	\end{subfigure}%
	\begin{subfigure}{.5\textwidth}
		\centering
		\includegraphics[width=.8\linewidth]{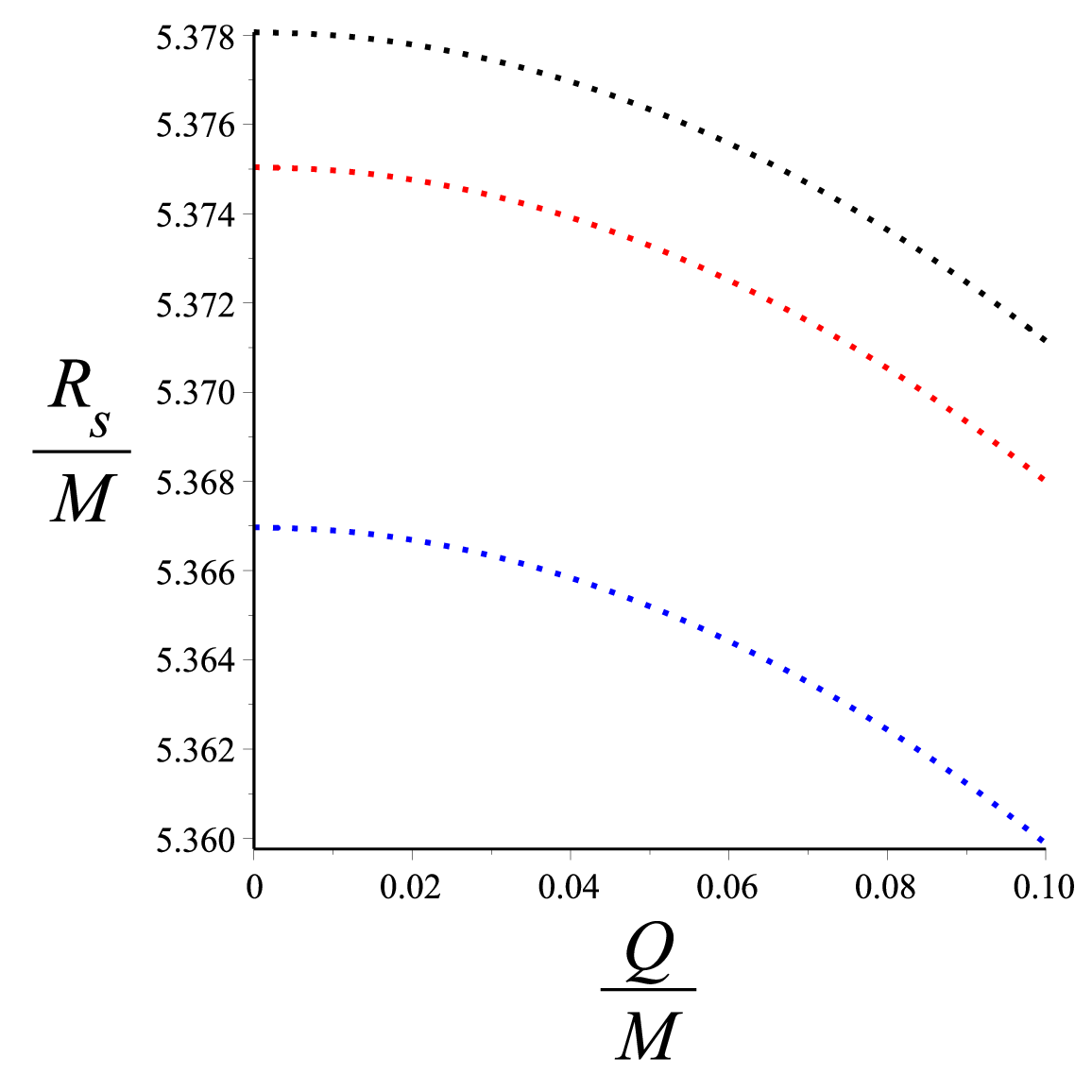}
		\caption{}
		\label{fig:Rsl05}
	\end{subfigure}
	\caption{Here we provide the numerical evaluations for the radius $R_s$. The black, red, and blue curves correspond to the cases of $C=-1$, $C=0$, and $C=1$, respectively. In Fig. \ref{fig:Rsl01}, we consider $l=0.1~M$ and $a=0.5~M$, whereas in Fig. \ref{fig:Rsl05} we have $l=0.5~M$ and $a=0.5~M$.}
\end{figure}

\begin{figure}[H]
	\centering
	\begin{subfigure}{.5\textwidth}
		\centering
		\includegraphics[width=.8\linewidth]{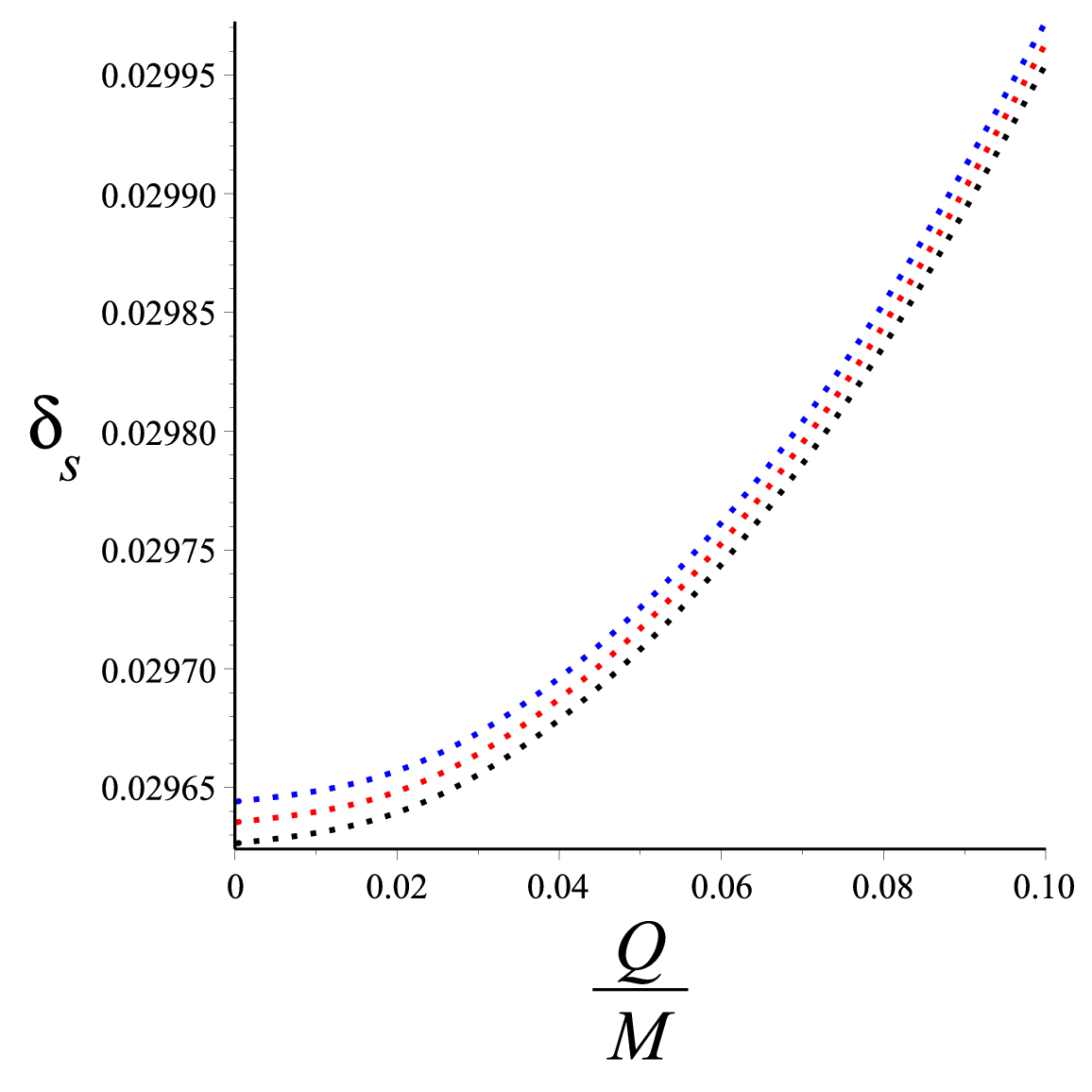}
		\caption{}
		\label{fig:dsl01}
	\end{subfigure}%
	\begin{subfigure}{.5\textwidth}
		\centering
		\includegraphics[width=.8\linewidth]{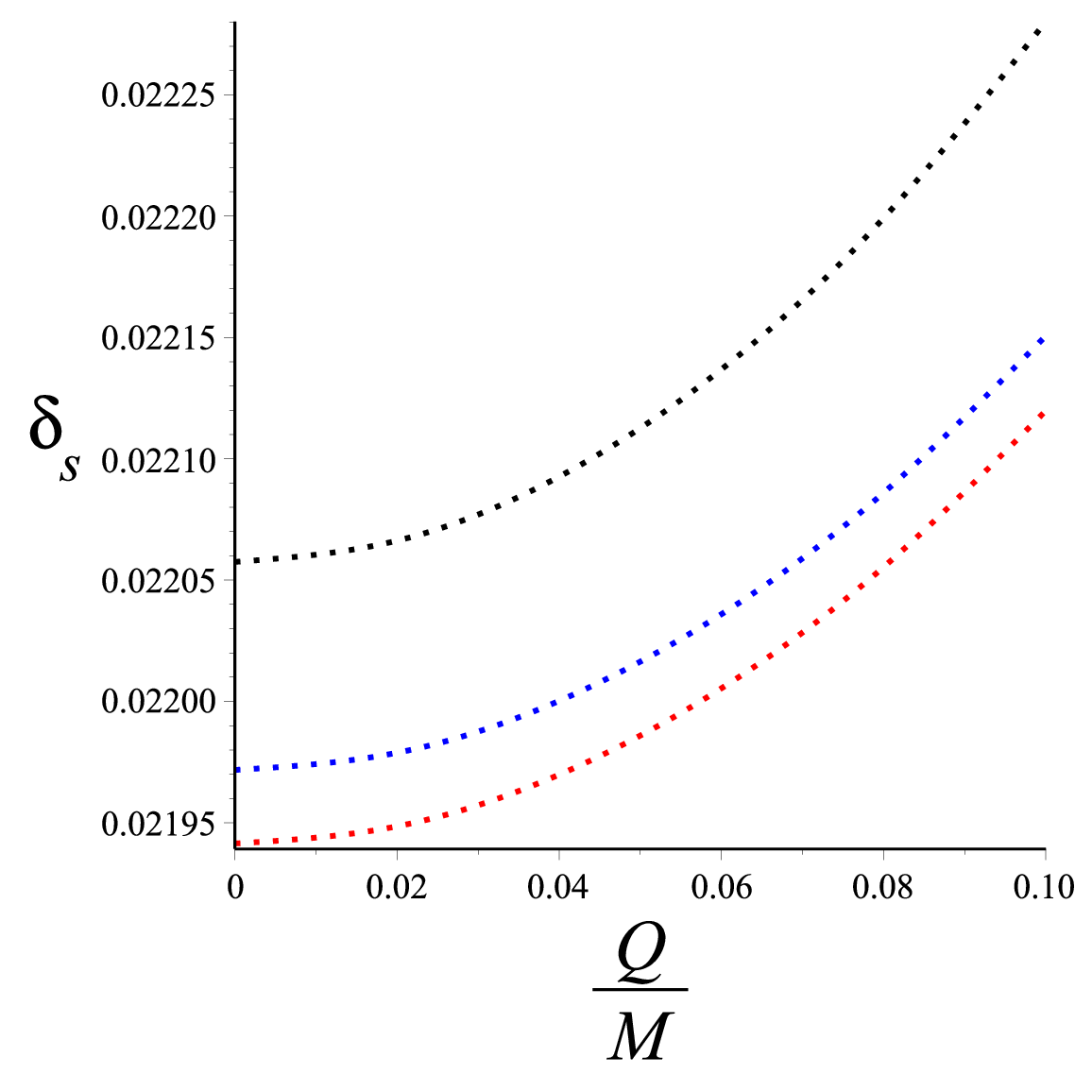}
		\caption{}
		\label{fig:dsl05}
	\end{subfigure}
	\caption{Here we provide the numerical evaluations for the radius $\delta_s$. The black, red, and blue curves correspond to the cases of $C=-1$, $C=0$, and $C=1$, respectively. In Fig. \ref{fig:dsl01}, we consider $l=0.1~M$ and $a=0.5~M$, whereas in Fig. \ref{fig:dsl05} we have $l=0.5~M$ and $a=0.5~M$.}
\end{figure}

From Figs. \ref{fig:Rsl01} and \ref{fig:Rsl05}, we observe that the effective radius $R_s$ decreases as the black hole's electric charge $Q$ grows. Note that the dependence of $R_s$ and $\delta_s$ on NUT parameter $l$ in Kerr-Taub-NUT spacetime has been worked out in \cite{Abdujabbarov:2012bn}. To align with the constraints proposed in \cite{Ghasemi-Nodehi:2024bcv} based on the Sgr A* data, we set the NUT parameter $l$ to $l=0.5~M$. These figures also demonstrate the effect of the MR parameter on the effective radius in the spacetime, showing that $R_s$ is largest for $C=-1$ and smallest for $C=1$. Furthermore, the overall size of the effective radius increases with a larger NUT charge. On the other hand, from Figs. \ref{fig:dsl01} we understand that as the NUT parameter increases, the black hole shadows for the three different values of $C$ become more distinguishable with less overlap.

\section{Conclusion}\label{sec.conclusion}

In this paper, we present a novel solution in the low-energy limit of heterotic string theory, describing a charged rotating black hole with NUT and Manko-Ruiz parameters. This work builds on previous studies, particularly \cite{Siahaan:2019kbw}, which extends the findings of \cite{Galtsov:1994pd} that discussed rotating NUT black holes within the low-energy limit of heterotic string theory. For this spacetime, we demonstrate the separability of the Hamilton-Jacobi equation for both timelike and null objects. The latter is particularly important for black hole shadow studies, following the methodology outlined in \cite{Cunha:2016bpi} for the shadow's edge and \cite{Hioki:2009na} for the observables. To illustrate how the black hole shadow's image is modified by variations in the MR, NUT, and electric charge parameters, we provide numerical results in Section \ref{sec.Shadow}. Additionally, we present the shadow edge observed by an observer at various inclination angles. In general, the behavior of the KSTN black hole shadow presented here is consistent with the shadow of NUT black holes in vacuum Einstein spacetimes, particularly in terms of the shift in the image center due to changes in the MR parameter \cite{Zhang:2021pvx} and the size of the observables as the NUT charge increases. For future work, we plan to investigate gravitational lensing in the KSTN spacetime. This study will consider both strong lensing, as discussed in \cite{Bozza:2002zj}, and weak lensing, as presented in \cite{Li:2020wvn}. Additionally, the phenomenon of superradiance in the KSTN spacetime presents an intriguing avenue for further exploration, as suggested in \cite{Lee:2023jfi}.
		
\section*{Acknowledgement}
		
This research is supported by LPPM UNPAR. I sincerely appreciate the constructive feedback provided by the anonymous referee.

\end{document}